\documentclass[12pt]{JHEP}
\usepackage{amsmath,amsfonts,amssymb}
 \usepackage{epsf}

\newcommand{\be}{\begin{equation}}
\newcommand{\ee}{\end{equation}}
\newcommand{\bea}{\begin{eqnarray}}
\newcommand{\eea}{\end{eqnarray}}

\newcommand{\tr}{\hbox{ Tr}}

\author{Nelia Mann$^{1,\dagger}$ and Samuel E. V\'azquez$^{2,\S}$\\
$^1$ Department of Physics, University of Chicago, 5640 S. Ellis
Ave., Chicago, IL 60637
 \\ $^2$
Department of Physics, University of California , Santa Barbara, CA
93106
\\
$^\dagger$\email{nelia@theory.uchicago.edu},
$^\S$\email{svazquez@physics.ucsb.edu}}

\title{Classical Open String Integrability}
\abstract{We present a simple procedure to construct  non-local
conserved charges for classical open strings on coset spaces. This
is done by including suitable reflection matrices on the classical
transfer matrix. The reflection matrices must obey certain
conditions for the charges to be conserved and in involution. We
then study bosonic open strings on $AdS_5\times S^5$. We consider
boundary conditions corresponding to Giant Gravitons on $S^5$,
$AdS_4\times S^2$ D5-branes and $AdS_5 \times S^3$ D7-branes. We
find that we can construct the conserved charges for the full
bosonic string on a Maximal Giant Graviton or a D7-brane. For the
D5-brane, we find that this is possible only in a $SU(2)$ sub-sector
of the open string. Moreover, the charges can not be constructed at
all for non-maximal Giant Gravitons.  We discuss the interpretation
of these results in terms of the dual gauge theory spin chains.}

\begin{document}

\section{Introduction}

In recent years, a great deal of attention  has been focussed on the proposed duality between String Theory on an $AdS_{5} \times S^{5}$ background and a $\mathcal{N} = 4$ supersymmetric Yang-Mills theory living on the boundary of this space \cite{ADSCFT}.  These theories, while seemingly very different, have many properties in common, such as global symmetries.  Direct comparison of general states in the two theories is made difficult by the weak/strong coupling nature of the duality, and for a long time, could only be applied to special states protected by supersymmetry.  More recently, we have been able to apply a variety of tools to study more general states.

In particular, there has been much discussion  of integrability on both sides of the dualtiy in the strict large $N$ limit.  Investigation of integrability in the gauge theory began when Minahan and Zarembo demonstrated that the one loop anomalous dimension operator, acting on single trace scalar operators, could be interpreted as the Hamiltonian of an integrable spin chain \cite{MZ}. Therefore, the anomalous dimensions could be found using a Bethe ansatz.  This work was quickly extended to the full set of single trace operators at one loop, and has been gradually extended beyond one loop.  For some of the work done on this, see \cite{spinchains}.  At present, no gauge theory calculation of these single trace operators has contradicted integrability.  Moreover, calculations in the gauge theory up to four loops have been shown to be consistent with integrability.

In the string theory, investigations began  with  the discovery of a
complete set of classically conserved non-local charges by Bena,
Polchinski, and Roiban \cite{BPR}; further exploration of these
charges can be found in \cite{charges}.  Now, a great deal of effort
has been put into using the tools of integrability  on both sides of
the duality to find and compare states.  For example, operators with
large R-charge that can be explored on the gauge theory side using
the Bethe ansatz, are dual to semi-classical strings, allowing for
direct numerical comparisons \cite{compare}.  The complete set of
non-local charges can be used to generate all such states as
solutions to integral equations, which can be compared with those
found on the gauge theory side \cite{general}.  But of course, the
evidence of the structure of integrability itself, on both sides of
the duality, is a compelling new argument in favor of the duality.

Another avenue of exploration  has been to discuss whether or not the integrability of these systems survives various perturbations and extends to other types of operators or strings.  Of particular interest to us has been to explore open strings ending on various types of D-branes embedded in the $AdS_{5} \times S^{5}$ space. These are dual to operators that can, in general, be treated as open spin chains.  Although there are other possibilities, in this paper we will discuss three such situations.  In the first, we wrap a D3-brane around an $S^{3}$ inside the $S^{5}$, creating a giant graviton, and discuss the integrability of open strings ending on it.  These giant gravitons appear in the $\mathcal{N} = 4$ gauge theory as baryonic operators, and open strings ending on them are dual to hybrid operators formed from combining a baryonic-type of operator with a ``word'', a string of fields like those in single-trace operators \cite{vijay}. In \cite{BV, bcv} it was shown that these operators can be modeled as open spin chains with the word forming the bulk of the chain, and the baryonic part of the operator creating the boundary interactions.  Such scalar open spin chains are known to be integrable at one loop for special, ``maximal'' giant gravitons. However, for non-maximal giant gravitons they cannot be solved using a Bethe ansatz, though limited evidence for some other form of integrability was found.

Another system  we will explore is created when a D5-brane is wrapped around an $AdS_{4}$ inside $AdS_{5}$ and an $S^{2}$ inside the $S^{5}$.  In this case, we must add an $\mathcal{N} = 2$ hypermultiplet of fundamental matter to the $\mathcal{N} = 4$ theory, and we confine this extra matter to a defect in the 3+1-dimensional space-time.  In this case, open strings ending on the D-brane are dual to operators where a ``word'' of adjoint matter is sandwiched between two fundamental fields confined to the defect. Again, such operators can be thought of as open spin chains, and at one loop and in the scalar sector, they are known to be integrable \cite{DM}.

Finally, we consider a D7-brane that fills $AdS_5$ and wraps an $S^2$ inside the $S^5$. The gauge theory dual is also created by adding a ${\cal N} = 2$ hypermultiplet to ${\cal N} = 4$ SYM.  This description is only valid in the probe-brane limit, where we keep the number of D7-branes added finite while sending $N$, the number of D3-branes, to infinity.  Outside of this limit, the branes back-react on the space and destroy the conformal symmetry.  An open string on this D-brane is also described in the gauge theory, at one-loop, by an integrable spin chain with open boundary conditions \cite{Erler}. For further work on the integrability of these three types of open spin chains and their dual open strings, see \cite{OSS}-\cite{2loopok}.

Our goal is to examine, for these three systems,  whether or not an
analog of the set of charges generated in \cite{BPR} can be found.
In order to do this, we will start by discussing how the boundary
conditions of putting the $1+1$-dimensional system on a finite  line
affect the construction of the charges.  In the past, charges have
been constructed for systems on a semi-infinite line, and it turns
out to be straightforward to generalize this to the finite line.
This will be demonstrated in section 2.  Another important aspect of
integrability is to show that the classically conserved charges are,
in fact, in involution.  This was done for the closed string in
$AdS_{5} \times S^{5}$ in \cite{nick1}, and we will show the finite
line variation of this result in section 3.  We will find that the
existence and involution of these charges then depend on the
specific boundary conditions used for the $1+1$ dimensional system,
which will vary depending on what types of D-branes we attach our
open strings to.

In section 4, we will study an $SU(2)$  sector of open strings
ending on the giant gravitons.  We will show that with these
boundary conditions, the classical charges can be found when the
giant gravitons are maximal, but that the technique fails for
non-maximal giant gravitons.  We will also discuss how this failure
relates to the problems with using a Bethe ansatz for the dual
operators.  In section 5 we will extend the analysis of maximal
giant gravitons to the full bosonic sector, again verifying
classical integrability. In section 6 we will switch to the D5-brane
system, and show that while restricted to an $SU(2)$ sector the
charges do exist, they do not exist for the full bosonic sector. In
section 6 we consider the case of the D7-brane. We show that the
full bosonic sector is integrable. In section 7 we will discuss our
conclusions and the open questions we believe still exist.

\section{Constructing Conserved Charges for Open Strings}

Here we will amend the techniques  that have  been used to find a
family of conserved charges for $1+1$ dimensional systems on a
periodic spatial dimension and for $1+1$ dimensional systems on a
semi-infinite line to show a technique that produces similar charges
on a finite line\footnote{The literature on $1+1$ field theories on
the half line is quite vast. The procedure used in this paper to
calculate the non-local charges, was first developed in
\cite{Bowcock} in the context of Affine Toda field theories. It was
latter applied for the $O(N)$ sigma model in \cite{Corrigan}. The
book \cite{book} contains a nice review and many references.}.

This is needed to study the integrability of open strings.   We will
assume we are working in the bosonic sector of the worldsheet action
for strings in $AdS_{5} \times S^{5}$.  However, the essential
argument does not depend on these details and could be applied quite
generally to integrable systems on a finite line.

The bosonic part of the worldsheet action for a string  (either open or closed) in $AdS_{5} \times S^{5}$ takes the form of the Principal Chiral Model. It is written in terms of the current $j = g^{-1}dg$ as
\begin{equation}
\label{action_basic} S = -\frac{\sqrt{\lambda}}{8\pi}\int \hbox{Tr}
(j \wedge \ast j)\;,
\end{equation}
where $g$ is an element of the coset $\frac{SO(4,2)}{SO(4,1)} \times \frac{SO(6)}{SO(5)}$. This current satisfies
\begin{equation}
d \ast j = 0\;,
\end{equation}
and
\begin{equation}
dj - j \wedge j = 0\;.
\end{equation}
One must also impose the Virasoro constraints:
\begin{equation}
 \hbox{Tr}(j_{\alpha} j_{\beta})  - \frac{1}{2} g_{\alpha \beta} g^{\gamma \delta}  \hbox{Tr}(j_{\gamma} j_{\delta})
  = 0\;.
 \end{equation}

Integrability generally hinges on the ability to create a one-parameter family of flat currents $J(x)$ such that
\begin{equation}
dJ - J \wedge J = 0\;,
\end{equation}
which we can do by using
\begin{equation}
J(x) = \frac{1}{1 - x^2}\left[j + x\ast j\right]\;.
\end{equation}

From them, we can construct an infinite set of conserved charges using the usual Monodromy matrix:
\begin{equation}
\label{opentrans} \Omega(\sigma_2, \sigma_1 ; x)  = P
\overleftarrow{\exp} \left(  \int_{\sigma_1}^{\sigma_2} d\sigma
J_1(\sigma; x) \right) \;.
\end{equation}
We have the basic properties,
\begin{eqnarray}
 &&\Omega(\sigma, \sigma; x)  = 1\;, \\
&&\Omega(\sigma_3, \sigma_2; x) \Omega(\sigma_2, \sigma_1; x)  = \Omega(\sigma_3, \sigma_1; x) \;, \\
&&  \Omega(\sigma_1, \sigma_2; x)^{-1} = \Omega(\sigma_2, \sigma_1; x) \;.
\end{eqnarray}
Moreover,
\begin{eqnarray}
 \partial_{\sigma_1} \Omega(\sigma_1, \sigma_2; x)  &=&  J_1(\sigma_1;x) \Omega(\sigma_1, \sigma_2; x) \;, \\
\partial_{\sigma_2} \Omega(\sigma_1, \sigma_2; x) & =&  -  \Omega(\sigma_1, \sigma_2; x) J_1(\sigma_2;x)\;, \\
\delta\Omega(\sigma_2, \sigma_1; x)   &=&
\int_{\sigma_1}^{\sigma_2} d\sigma \Omega(\sigma_2, \sigma; x)
\delta J_1(\sigma;x) \Omega(\sigma, \sigma_1; x) \;.
\end{eqnarray}

Using these last relations along with the flatness of $J$ it is easy to prove that,
\begin{equation}
 \partial_\tau \Omega(\sigma_2, \sigma_1;
x) =  -\Omega(\sigma_2, \sigma_1; x)J_0(\sigma_1;x) +J_0(\sigma_2;x)
\Omega(\sigma_2, \sigma_1 x)\;.
\end{equation}
Therefore, under the periodic boundary conditions of the closed string we have an infinite family of conserved charges,
 \begin{equation}
  \partial_\tau\hbox{Tr}\Omega_{c}(x)^n = 0\;,
\end{equation}
 where $\Omega_{c}(x) \equiv \Omega(2\pi, 0; x)$.

Using as inspiration both the techniques  of open spin chains, their relationships to closed spin chains and the methods for producing charges on a semi-infinite line, we suppose that the correct object to substitute for $\Omega_{c}(x)$ would involve an integral that is taken over the open string from one end to the other, and then back the other direction.  We include matrices $\kappa_{0, \pi}$ which represent reflection off the ends of the string.  Thus we have the objects
\begin{equation}
\label{Omega} \Omega(x) \equiv \kappa_0(x) \Omega_R(2\pi ,\pi ; x)
\kappa_\pi(x) \Omega(\pi ,0 ;x)  \;,
\end{equation}
 where $\Omega_R$ is constructed with the reflected value of the fields:
 \begin{equation}
 (j_{R})_{0}(\sigma) = j_{0}(2\pi - \sigma), \ \ \ \ \ (j_{R})_{1}(\sigma) = -j_{1}(2\pi
-\sigma),
 \end{equation}
and $\sigma \in [\pi , 2\pi]$. We find that the charges $\Omega(x)$ will satisfy $\partial_{\tau} \hbox{Tr}\Omega(x)^{n} = 0$ when the matrices $\kappa_{0, \pi}$ satisfy the conditions
\begin{eqnarray}
 \label{reflection}\partial_\tau \kappa_0(x)- J_0( 0;x)
\kappa_0(x)+ \kappa_0(x) (J_R)_0( 0 ; x) &=& 0\;, \\
\partial_\tau \kappa_\pi(x) - (J_R)_0( 0;x)
\kappa_\pi(x) +\kappa_\pi(x) J_0( 0 ; x) &=& 0\;.
\end{eqnarray}

Given some asymptotic value of the reflection matrices, say at $\tau = -\infty$,  the most general solution to the above equations is,

\begin{eqnarray}
 \kappa_0(\tau;x) &=& P\overleftarrow\exp\left(
\int_{-\infty}^\tau d\tau' J_0(\tau', 0;x) \right)
\kappa_0(-\infty;x) P\overleftarrow\exp\left( \int^{-\infty}_\tau
d\tau' (J_R)_0(\tau', 0;x) \right)\;, \nonumber \\
\kappa_\pi(\tau;x) &=& P\overleftarrow\exp\left( \int_{-\infty}^\tau
d\tau' (J_R)_0(\tau', \pi;x) \right) \kappa_\pi(-\infty;x)
P\overleftarrow\exp\left( \int^{-\infty}_\tau d\tau' J_0(\tau',
\pi;x) \right)\;. \nonumber
\\
\end{eqnarray}

However, this solution is not  acceptable because it is non-local in time and it would not be possible to compute Poisson brackets with these reflection matrices. Therefore, we conclude that the reflection matrices must be time independent. This is a  significant constraint since, even if we set $\partial_\tau \kappa = 0$ in (\ref{reflection}), we are not guaranteed that the solution is time independent since, in general, the matrices $J_0$ and $(J_R)_0$ will depend on time.

Using the conformal gauge for the world-sheet metric (which we will maintain for the duration of this paper), the condition on the reflection matrices becomes
\begin{equation}
 \label{refcg}
[j_0 ,\kappa_{0 ,\pi} ]  \pm x \{j_1, \kappa_{0,\pi}\} = 0\;,
\end{equation}
where for $\kappa_0$ we have a $+$ sign  and the currents are evaluated at $\sigma = 0$, while for $\kappa_{\pi}$ we have a $-$ sign and the currents are evaluated at $\sigma = \pi$. Therefore, we can look, without loss of generality, to the boundary at $\sigma = 0$ and get the other reflection matrix by inverting the sign of $x$. For the remainder of this paper we will do this, dropping the subscript notation $\kappa_{0} \rightarrow \kappa$ and assuming a plus sign in this equation.  The question of whether or not open strings with a particular set of boundary conditions (determined by the configuration of the D-brane on which they end) have this complete set of conserved charges, can be answered by exploring whether or not a time-independent solution to equation (\ref{refcg}) can be found.

\section{Canonical Structure}

We can also explore whether or not the conserved charges found above are in involution. We will only check this explicitly for open strings in the  $SU(2)$ sector.  However, for completeness, we present the general procedure. We will use the conformal gauge from now on.

First, we need to find the canonical structure of the model at hand. For the Principal Chiral Model, it is well known that Poisson brackets of the current $J_1$ can be written in the $r-s$ formalism introduced by Maillet \cite{Maillet1, Maillet2, Maillet3},
\bea
\label{J1bracket} \{ J_1(\sigma, x) \overset{\otimes}{,}
J_1(\sigma',x')\} &=&
r'(\sigma, x,x') \delta(\sigma - \sigma') \nonumber\\
&& + [r(\sigma,x,x'), J_1(\sigma,x) \otimes \mathbf{1} + \mathbf{1}
\otimes J_1(\sigma',x')]\delta(\sigma - \sigma')\nonumber\\
&&- [s(\sigma,x,x'), J_1(\sigma ,x) \otimes \mathbf{1} -
\mathbf{1}\otimes J_1(\sigma',x')]\delta(\sigma-\sigma')\nonumber\\
&& - [s(\sigma,x,x') + s(\sigma',x,x')] \delta'(\sigma -
\sigma')\;.
\eea
The explicit form of the functions $r$ and $s$ depends on the model, but they only depend on the current $j$ and not on its derivatives. The terms with derivatives on the delta function are  called ``Non-Ultra-Local" (NUL).

Using the properties of the transfer matrix, one can easily show that
\begin{eqnarray}
\label{Omegabracket}
\{ \Omega(\sigma_1, \sigma_2; x) \overset{\otimes}{,}  \Omega(\sigma_1', \sigma_2'; x') \} &=& \int_{\sigma_2}^{\sigma_1} d\sigma  \int_{\sigma_2'}^{\sigma_1'} d\sigma'  \left( \Omega(\sigma_1, \sigma; x)\otimes \Omega(\sigma_1', \sigma'; x') \right) \nonumber \\
&& \times \{J_1(\sigma; x)\overset{\otimes}{,} J_1(\sigma'; x)\}
 \left( \Omega(\sigma, \sigma_2; x)\otimes \Omega(\sigma', \sigma_2'; x') \right)\;. \nonumber \\
 \end{eqnarray}
It is well known that the NUL terms in (\ref{J1bracket}) produce a discontinuity in (\ref{Omegabracket}) when any of the end points $\sigma_i, \sigma_i'$ coincide.  The correct brackets are defined by the Maillet's regularization procedure \cite{Maillet1,Maillet2,Maillet3}.  We will not go into the details of this which is reviewed in \cite{nick1}. Here we will only need the following result \cite{Maillet1} (see also \cite{book}),
\bea
\label{Omegareg} \{ \Omega(\sigma_1,\sigma_2;x)\overset{\otimes}{,}
\Omega(\sigma_1,\sigma_2;x')\} &=& \epsilon(\sigma_1 - \sigma_2)
\left[ r(\sigma_1,x,x') \Omega(\sigma_1,\sigma_2;x) \otimes
\Omega(\sigma_1,\sigma_2;x') \right. \nonumber \\
&&\left.- \Omega(\sigma_1,\sigma_2;x) \otimes
\Omega(\sigma_1,\sigma_2;x') r(\sigma_2,x,x')\right]\;,\eea were
$\epsilon(\sigma)  = \text{sign}(\sigma)$. The consistency of the
Poisson Brackets, imply that \bea\label{step} 0 &=&
\{\Omega(\sigma_1,\sigma_2;x)\overset{\otimes}{,}
\Omega(\sigma_1,\sigma_2;x')\Omega(\sigma_2,\sigma_1;x')\}\nonumber
\\
& =& (\mathbf{1}\otimes \Omega(\sigma_1,\sigma_2;x'))
\{\Omega(\sigma_1,\sigma_2;x)\overset{\otimes}{,}
\Omega(\sigma_2,\sigma_1;x')\} \nonumber \\
&&+ \{\Omega(\sigma_1,\sigma_2;x)\overset{\otimes}{,}
\Omega(\sigma_1,\sigma_2;x')\} (\mathbf{1}\otimes
\Omega(\sigma_2,\sigma_1;x')).\eea Using (\ref{Omegareg}) and
(\ref{step}) we get, \bea \label{Omegareg1}&&\{ \Omega(\sigma_1,
\sigma_2; x) \overset{\otimes}{,}
\Omega^{-1}(\sigma_1, \sigma_2; x') \} \nonumber \\
& =& \epsilon(\sigma_1 - \sigma_2) \left[\Omega(\sigma_1,\sigma_2;
x) \otimes\mathbf{1}\right) r(\sigma_2, x,x')
\left(\mathbf{1}\otimes \Omega^{-1}(\sigma_1,\sigma_2;x')\right)\nonumber \\
 &&  -\left(\mathbf{1}\otimes \Omega^{-1}(\sigma_1,\sigma_2;x')\right) r(\sigma_1,x,x')
\left( \Omega(\sigma_1,\sigma_2; x) \otimes\mathbf{1}\right]\;.
\eea

It is easy to show that, in conformal gauge, $\Omega_R(2\pi, \pi ; x) = \Omega(0,\pi;-x)$. Thus, \be\label{Omegaconf} \Omega(x) \equiv \kappa_0(x) \Omega^{-1}(\pi ,0 ;- x) \kappa_\pi(x) \Omega(\pi ,0 ;x) \;.\ee Let us now assume that the reflection matrices  do not depend on the fields or their derivatives.  This is not completely general, but will suffice for the cases examined in this paper.  Using (\ref{Omegareg}) and (\ref{Omegareg1}), it is a matter of algebra to compute the Poisson brackets of the transfer matrices (\ref{Omegaconf}).

The result is that,
\begin{equation}
\{ \tr \Omega(x), \tr \Omega(x') \} = 0\;,
\end{equation}
provided that the  reflection matrices obey,
\begin{eqnarray}
\label{conv} [ r(0,x,x') , \kappa_{0}(x)\otimes \kappa_0(x') ] &&+
(\kappa_0(x)\otimes\mathbf{1} )r(0,x,-x')
(\mathbf{1}\otimes \kappa_0(x'))\nonumber \\
&& - (\mathbf{1}\otimes \kappa_0(x'))
r(0,x,-x')(\kappa_0(x)\otimes\mathbf{1} ) = 0\;,
\end{eqnarray}
with a similar equation for the other boundary. This was the condition found in \cite{Corrigan} for the case of the $O(N)$ model on the half line. Here we see that it also applies in the case of two boundaries.

If the current $j$  takes value in the Lie algebra $\mathbf{su}(2)$,
it was shown in \cite{nick1} that $r(\sigma,x,x')$ takes the simple
form, \bea\label{rsu2} r(x,x') &=&   \frac{2\pi}{\sqrt \lambda}
\frac{x^2 + x'^2 - 2 x^2 x'^2}{(x- x')(1 - x^2)(1 - x'^2)}  (t^a
\otimes t^a)\;, \eea where $t^a$ are the $\mathbf{su}(2)$ generators
normalized as $\tr(t^a t^b) = - \delta^{a b}$. We will come back to
this case in the next sections.

\section{An $SU(2)$ Sector of Open Strings Ending on Giant Gravitons}

The above analysis gives us a quite general  method for exploring whether or not open strings with a particular set of boundary conditions are integrable.  We would now like to apply this method to a few different cases; this will serve to illustrate the technique as well as giving us insight into these systems. The first case we will consider is that of open strings ending on giant gravitons in $AdS_{5} \times S^{5}$.  The giant gravitons (GG)  we will be considering with wrap an $S^{3}$ inside the $S^{5}$. They were first explored in \cite{GG}, and open strings ending on them have been studied in \cite{ bcv}.  Both the GGs themselves and the open strings ending on them, can be understood in the dual $\mathcal{N} = 4$ super Yang-Mills theory as a particular type of operator \cite{vijay}, and the integrability of these operators was explored in \cite{BV}.  If we write the metric of $AdS_{5} \times S^{5}$ in terms of global coordinates
\begin{equation}
ds^2 = -\cosh^2\rho dt^2 + d\rho^2 + \sinh^2\rho d\Omega_{3}'^2 +
\sin^2\theta d\phi^2 + d\theta^2 + \cos^2\theta d\Omega_{3}^2\;,
\end{equation}
then we locate the D3-brane  that is the giant graviton at $\rho = 0$ and $\theta = \theta_{0}$, so that it will wrap the $\Omega_{3}$ 3-sphere.  The giant graviton then carries angular momentum along the $\phi$ direction, so $\phi = \phi(\tau)$; in fact, we set $\phi = t$.  Using
\begin{equation}
d\Omega_{3}^2 = \cos^2\psi d\varphi^2 + d\psi^2 + \sin^2\psi d\eta^2,
\end{equation}
the coordinates $\varphi$, $\psi$, $\eta$, and $t$ parametrize the D-brane. The maximal giant graviton (MGG) has $\theta_0 = 0$.

To simplify matters we will only consider an $SU(2)$  sector to begin with. Here we will find it easiest to amend the form of the action (\ref{action_basic}) to read
\begin{equation}
S = -\frac{\sqrt{\lambda}}{4\pi}\int \left[\frac{1}{2}\hbox{Tr}(j
\wedge \ast j) + dt \wedge \ast dt\right]\;,
\end{equation}
where $j = -g^{-1}dg$ and
\begin{equation}
g = \left(\begin{array}{cc} Z & Y \\ -\bar{Y} & \bar{Z}
\end{array}\right) \in SU(2) \;,
\end{equation}
where $|Z|^2 + |Y|^2 = 1$.  In terms of the coordinates of the 5-sphere given above, we have
\begin{equation}
\label{embedding} Z = \sin\theta e^{i\phi}, \ \ \ \ \ Y = \cos\theta
e^{i\varphi}.
\end{equation}
(Choosing the $SU(2)$ sector involves restricting the string  to $\rho = 0$ and $\psi = 0$ everywhere on the string.)

We will also need to use the boundary conditions for these strings.  Keeping in mind that we need $\delta\phi = \delta t$ at the boundary, variation of the action gives us
\begin{equation}
\int d\tau \sqrt{-g} g^{\sigma \alpha}\left[\delta t (-\partial_{\alpha} t  + \sin^2\theta_0 \partial_{\alpha} \phi) + \delta \varphi \cos^2\theta_0 \partial_{\alpha}\varphi \right] |_{\sigma = 0}^{\sigma = \pi} = 0\;.
\end{equation}

Therefore, at each boundary we need,
\begin{eqnarray}
 \label{tbc} (-t'  + \sin^2\theta_0 \phi')   &=& 0\;,\\
  \varphi' &=& 0\;.
\end{eqnarray}
We are  using the notation that $f'$ and $\dot{f}$ stand for derivatives of the function $f(\tau, \sigma)$ with respect to $\sigma$ and $\tau$ respectively.  Note that, as opposed to the case of the closed string, the EOM of $t$ is not completely decoupled from the rest of the coordinates. Therefore, in general we do not have the simple solution $t \sim \tau$.

Now, lets consider what condition (\ref{refcg}) will give us.  In the SU(2) sector we can simplify matters by expressing the current $j$ and the reflection matrix $\kappa$ as linear combinations of sigma matrices and the $2 \times 2$ identity matrix:
\begin{equation}
j_{\alpha} = j_{\alpha}^{a}\sigma^{a}, \ \ \ \ \ \ \kappa = \kappa^{0}I + \kappa^{a}\sigma^{a}
\end{equation}
in which case we can use the familiar identities $[\sigma^{a}, \sigma^{b}] = 2i\epsilon^{abc}\sigma^{c}$ and $\{\sigma^{a}, \sigma^{b}\} = 2\delta^{ab}I$ to rewrite (\ref{refcg}) as
\begin{equation}
\left(\begin{array}{cccc} 0 & xj_{1}^{1} & xj_{1}^{2} & xj_{1}^{3} \\ xj_{1}^{1} & 0 & -ij_{0}^{3} & ij_{0}^{2} \\ xj_{1}^{2} & ij_{0}^{3} & 0 & -ij_{0}^{1} \\ xj_{1}^{3} & -ij_{0}^{2} & ij_{0}^{1} & 0 \end{array}\right)\left(\begin{array}{c} \kappa^{0} \\ \kappa^{1} \\ \kappa^{2} \\ \kappa^{3} \end{array}\right) \equiv \mathbf{M}\vec{\kappa} = 0.
\end{equation}
It is clear that in order for this equation to have a non-trivial solution we need $ \det \mathbf{M} = 0$.  It is straightforward to demonstrate that this implies $\hbox{Tr}(j_{0}j_{1}) = 0$. In addition, because we need $\dot{\kappa} = 0$, we also must have $\det \dot{\mathbf{M}} = 0$. This gives $\hbox{Tr}(\partial_{0}{j}_{0}\partial_{0}{j}_{1}) = 0$.  The first of these conditions implies that
\begin{equation}
\label{cond1} \sin^2\theta_{0} \, \dot{\phi} \, \phi' = 0.
\end{equation}
This condition is always satisfied for the MGG.

In this case, one can show that the only possible time independent reflection matrix is given by,
\be \kappa_\text{MGG}
\propto \sigma^{3}\;.\ee Using \be [\sigma^a \otimes \sigma^a,
\sigma^3 \otimes \sigma^3] = 0\;,\;\;\; \sigma^3 \sigma^a \otimes
\sigma^a \sigma^3 - \sigma^a \sigma^3 \otimes \sigma^3 \sigma^a =
0\;,
\ee
one can show that the condition (\ref{conv}) is indeed satisfied. Therefore, the conserved charges are in involution for an $SU(2)$ open string on a MGG.

On the other hand, for non-maximal GGs ($\theta_{0} \ne 0$), one clearly has an extra condition on the string at the boundaries: $\phi' = 0$. Thus, open strings ending on giant gravitons will not, in general, be proven integrable by this method. Even assuming assuming $\phi' = 0$ at the boundary, $\det \do {\mathbf{M}} = 0$ gives an additional condition:
\begin{equation}
\label{cond2} \cos\theta_{0}\sin\theta_{0}(\dot{\varphi} -
\dot{\phi})\left[\theta'(\ddot{\varphi} + \ddot{\phi}) -
\dot{\theta}'(\dot{\varphi} + \dot{\phi})\right] = 0.
\end{equation}
This condition is not implied by either the EOMs or the Virasoro constraints.

The condition $\phi' = 0$ at the string boundary has an interesting physical meaning.  We know that for the closed string there are conserved charges associated with left- and right-multiplication of an element of $SU(2)$.  These would be
 \begin{equation} Q_R = \frac{\sqrt \lambda}{4 \pi} \int_\gamma
\ast j\;, \;\;\;\; Q_L = \frac{\sqrt \lambda}{4 \pi} \int_\gamma
\ast( g j g^{-1} )\;,
\end{equation}
where $\gamma$ is any closed curve winding once around the world-sheet.  Because they are conserved classically, we can focus on just the `highest weight' solutions with
 \begin{equation}
 Q_{R} = \frac{1}{2i}R\sigma^{3}, \ \ \ \ \ Q_{L} = \frac{1}{2i}L\sigma^{3}, \ \ \ \ R, L \in \mathcal{R}^{+}.
 \end{equation}
In contrast, the analogous charges are not both conserved for general open strings.  In this case, we have
\begin{equation}
\partial_\tau Q_R  =
\frac{\sqrt \lambda}{4\pi}(\ast j)_0|_{\sigma = 0}^{\sigma  =
\pi}\;,
\end{equation}
and
 \begin{equation}
  \partial_\tau Q_L  = \frac{\sqrt
\lambda}{4\pi}[\ast( g j g^{-1})]_0|_{\sigma = 0}^{\sigma  =
\pi}\;.
\end{equation}
Then, it is easy to show that
\begin{eqnarray}
 \partial_\tau(L + R) &=& - \frac{\sqrt\lambda}{\pi} \sin^2\theta_{0} \, \phi'|_{\sigma = 0}^{\sigma  = \pi} \;,\\
\partial_\tau(L - R) &=& 0\;.
\end{eqnarray}
The non-conservation  of $L+R$ for these open strings reflects the way that the open string is dragged behind the giant graviton and the two objects can exchange angular momentum \cite{bcv}.  In the gauge theory dual, it was shown that this requires that the dual open spin chain is not of fixed length and cannot be solved by a simple Bethe ansatz.  Note that in the special case of the maximal giant graviton $\theta_{0} = 0$ no such problem exists, as $L+R$ is conserved and the dual spin chains do not change length and can be solved using a Bethe ansatz (at least at one loop).

Now we see that the condition $\phi' = 0$ at  the boundary corresponds to requiring, by hand, that the quantity $R + L$ is conserved and thus that angular momentum is not being exchanged between the open string and the giant graviton.  The same feature that provides an obstacle to integrability in the dual open spin chains is standing in our way here.  However, the second condition (\ref{cond2}) demonstrates that $\phi' = 0$ by itself is not a good enough constraint to give us this construction of
charges.  We could solve the problem by adding the requirement $\phi = t = \pm \varphi$ at the boundaries, which corresponds to the endpoints of the string following null geodesics along the giant
graviton.  Alternatively, we could consider an ``extremal giant graviton'' with $\theta_{0} = \frac{\pi}{2}$ along with the boundary condition $\phi' = 0$.  But these are all extremely restrictive constraints, so it is clear that the general open string ending on a non-maximal giant graviton cannot be demonstrated to be integrable by this approach.   On the other hand, our failure to find a complete set of non-local conserved charges by this technique is not sufficient to demonstrate that these open strings are not integrable.

\section{The Full Bosonic Sector for Open Strings Ending on MGGs}

In the previous section, we showed that in the  $SU(2)$ sector, open strings ending on MGGs are classically integrable.  Here we would like to extend the analysis, just for the MGGs, to the full bosonic sector.

For the full bosonic model, we consider the coset \be AdS_5 \times
S^5 =  \frac{SO(4,2)}{SO(5,1)} \times \frac{SO(6)}{SO(5)} \;.\ee One
can embed and element $g_a$ of the coset $SO(4,2)/ SO(5,1)$ into the
group $SU(2,2)$ which is locally isomorphic to $SO(4,2)$. Similarly,
we can embed an element $g_s$ of $SO(6)/SO(5)$ into $SU(4)$ which is
isomorphic to $SO(6)$. The embedding is the following
\cite{Arutyunov}: \be \label{blocdiag}
 g = \begin{pmatrix} g_A & 0 \\
 0 & g_S\end{pmatrix}\;,
 \ee
where,
\be
\label{gmatrices}
g_A =\begin{pmatrix} 0 &  {\cal Z}_3 & - {\cal Z}_2 & \bar{ {\cal Z}_1}  \\
- {\cal Z}_3 & 0 &  {\cal Z}_1 & \bar{{\cal Z}_2} \\
{\cal Z}_2& - {\cal Z}_1 & 0 & - \bar{{\cal Z}_3}\\
-\bar{{\cal Z}_1} & - \bar{{\cal Z}_2} & \bar{{\cal Z}_3} & 0 \end{pmatrix}\;, \;\;\;\; g_S =\begin{pmatrix} 0 &  {\cal Y}_1 & - {\cal Y}_2 & \bar{ {\cal Y}_3}  \\
- {\cal Y}_1 & 0 &  {\cal Y}_3 & \bar{{\cal Y}_2} \\
{\cal Y}_2& - {\cal Y}_3 & 0 & \bar{{\cal Y}_1}\\
-\bar{{\cal Y}_3} & - \bar{{\cal Y}_2} & -\bar{{\cal Y}_1} & 0 \end{pmatrix}\;.
\ee
Since $g_A$ obeys $SU(2,2)$ it satisfies,
\be g_A^\dagger E g_A = E\;,\;\;\; E = \text{diag}(-1,-1,1,1)\;.\ee
Analogously, $g_S^\dagger g_S = \sum_i |{\cal Y}_i|^2 =  1$. Moreover,
\be -| {\cal Z}_3|^2 + |{\cal Z}_1|^2  + |{\cal Z}_2|^2 = -1\;.\ee

We can parametrize $AdS_5 \times S^{5}$ as,
\bea {\cal Z}_1 &=& \sinh \rho \xi_1\;,\;\;\; {\cal Z}_2 = \sinh \rho \xi_2\;,\;\;\; {\cal Z}_3 = \cosh \rho e^{i t}\;, \nonumber \\
{\cal Y}_2  &=& \cos \theta \Omega_2\;, \;\;\;  {\cal Y}_3  = \cos \theta \Omega_3\;,\;\;\; {\cal Y}_1 = \sin\theta e^{i \phi}\;,\eea
where $\sum_i |\xi_i|^2 = 1$ and $\sum_i |\Omega_i|^2 = 1$.

The boundary conditions for the MGG are: $\rho = 0, \theta =0 , \phi = t$, and Neumann in the other directions.  (Note that the complication in the boundary condition (\ref{tbc}) is removed by restricting to the maximal giant graviton.)

Given the form of the group element $g$, we can take the reflection matrices as,
\be\kappa = \begin{pmatrix}  \kappa_A &0 \\
0 &\kappa_S\end{pmatrix}\;.\ee
Therefore we can separate the problem in the $AdS_5$ and $S^5$ part. The $AdS$ currents $j_A$ at the boundaries are,
\bea (j_A)_0 &=&  i \partial_{0}t \; \text{diag}(1,1,-1,-1)\;,\\
(j_A)_1 &=&   i \rho'  \begin{pmatrix} 0 & A \\
-A^\dagger & 0 \end{pmatrix}\;,\eea
where,
\be A = i e^{it} \begin{pmatrix} \bar \xi_1 & \xi_2\\
\bar \xi_2& - \xi_1\end{pmatrix} \in SU(2)\;.\ee
Therefore it is very easy to see that, just like in the SU(2) case,   we can satisfy Eq. (\ref{refcg}) if we  take $\kappa_A$ along $(j_A)_0$. That is,
\be \kappa_A \propto \text{diag}(1,1,-1,-1)\;.\ee

The $S^5$ looks a little messier but also has a simple solution.  We have,
\be (j_S)_0  = \sum_{a = 1}^{3} \begin{pmatrix} \alpha^a \sigma^a & 0 \\
0 & \beta^a \sigma^a \end{pmatrix}\;,\ee
where
\bea
\alpha^1= \dot \Omega_2 \bar{\Omega}_2 + \Omega_3 \dot{ \bar \Omega}_3\;,\;\; \beta^1 = \dot \Omega_2 \bar \Omega_2+ \bar \Omega_3  \dot \Omega_3 \;,\\
\alpha^2 = - i \text{Re}(\Omega_2 \dot{ \bar \Omega}_3 - \bar \Omega_3 \dot \Omega_2)\;,\;\;\; \beta^2 =
- i \text{Re}(\Omega_2 \dot{ \Omega}_3 - \Omega_3 \dot \Omega_2)\;,\\
\alpha^3 =  i \text{Im}(\Omega_2 \dot{ \bar \Omega}_3 - \bar
\Omega_3 \dot \Omega_2)\;,\;\;\; \beta^3 =  i \text{Im} (\Omega_2
\dot{ \Omega}_3 - \Omega_3 \dot \Omega_2)\;.\eea Moreover,
\be (j_S)_1  = i \theta' \begin{pmatrix} 0 & B \\
- B^\dagger & 0 \end{pmatrix}\;,\ee
where,
\be B = i e^{i X_0} \begin{pmatrix} \bar \Omega_3 & \Omega_2 \\
\bar \Omega_2& - \Omega_3\end{pmatrix} \in SU(2)\;.\ee

Again, $\kappa_{S} \propto \text{diag}(1,1,-1,-1)$ will do the trick.   Clearly, the classical integrability of open strings ending on maximal giant gravitons is valid for the entire bosonic sector.

\section{The Bosonic sector for Open Strings Ending on D5-branes}

We can obtain a different set of boundary  conditions for our open strings by attaching them to D5-branes, arranged so that the brane wraps an $AdS_{4}$ inside the $AdS_{5}$, and an $S^{2}$ inside the $S^{5}$.  We will stay in the probe brane limit for this calculation, neglecting the back reaction of the D-branes on the spacetime.  However, the gauge theory results of one-loop integrability for the dual open spin chains made no distinction between this limit and the more general situation where the number of D5-branes is comparable to the number of D3-branes. We will study the bosonic sector of open strings attached to this brane using the same set up as in the previous section.  Therefore, the starting point is a matrix $g$ in the block diagonal form of (\ref{blocdiag}).

Now, the form of $g_{S}$ we want to use is the same as that given in (\ref{gmatrices}), except we will use angular coordinates defined as
\begin{equation}
{\cal Y}_{1} = X_{1} + iX_{2}, \ \ \ \ \ {\cal Y}_{2} = X_{3} +
iX_{4}, \ \ \ \ \ {\cal Y}_{3} = X_{5} + iX_{6}\;,
\end{equation}
with
\begin{eqnarray}
X_{1} & = & \cos\theta\cos\varphi\cos\eta\;, \nonumber \\
X_{2} & = & \cos\theta\cos\varphi\sin\eta\;, \nonumber \\
X_{3} & = & \cos\theta\sin\varphi\;, \nonumber \\
X_{4} & = & \sin\theta\sin\psi\;, \nonumber \\
X_{5} & = & \sin\theta\cos\psi\cos\xi \;,\nonumber \\
X_{6} & = & \sin\theta\cos\psi\sin\xi\;.
\end{eqnarray}

The brane is located at $\theta = 0$, so the variables $\varphi$ and $\eta$ satisfy Neumann boundary conditions, while $\theta$ satisfies Dirichlet boundary conditions.  Note that an $SU(2)$ sector is achieved by requiring $\varphi = \psi = 0$ along the whole string (so all derivatives of these variables are also zero).

We can calculate that
\begin{equation}
j_{S,0} = \left(\begin{array}{cc} f & g\sigma^{1} \\
-\bar{g}\sigma^{1} & -f \end{array}\right)\;,
\end{equation}
with
\begin{equation}
f = i\cos^2\varphi \, \dot{\eta}\;,
\end{equation}
and
\begin{equation}
g = e^{i\eta}(\dot{\varphi} - i\sin\varphi\cos\varphi \,
\dot{\eta})\;,
\end{equation}
 while
\begin{equation}
j_{S,1} = \theta' \left(\begin{array}{cc} \alpha^{a}\sigma^{a} &
\beta^{0}I + \beta^{a}\sigma^{a} \\ \gamma^{0}I +
\gamma^{a}\sigma^{a} & \delta^{a}\sigma^{a} \end{array}\right)\;.
\end{equation}

These variables are not all independent:
\begin{eqnarray}
\alpha^{3} = \delta^{3} & = & i\sin\varphi\sin\psi\;, \nonumber \\
\alpha^{2} = \delta^{2} & = & i\sin\varphi\cos\psi\cos\xi \;,\nonumber \\
\alpha^{1} = -\delta^{1} & = & -i\sin\varphi\cos\psi\sin\xi\;, \nonumber \\
\beta^{0} = -\bar{\gamma}^{0} & = & -i\cos\varphi\cos\psi\sin\xi e^{i\eta} \;,\nonumber \\
\beta^{3} = -\bar{\gamma}^{3} & = & \cos\varphi\cos\psi\cos\xi e^{i\eta} \;,\nonumber \\
\beta^{1} = \gamma^{1} & = & 0 \;,\nonumber \\
\beta^{2} = -\bar{\gamma}^{2} & = & -\cos\varphi\sin\psi
e^{i\eta}\;.
\end{eqnarray}
It is clear that if a general string ending on the  D5-brane is to be integrable, we must have $[j_{S, 0}, \kappa_{S}] = 0$ and $\{j_{S, 1}, \kappa_{S}\} = 0$ separately, because otherwise we can not have $\theta'$ and $\dot{\eta}$ and $\dot{\varphi}$ independent. (The EOM do not give any natural relations between these at the boundary.)  Considering the first condition, and making
\begin{equation}
\kappa_{S} = \left(\begin{array}{cc} A & B \\ C & D
\end{array}\right)\;,
\end{equation}
we find that
\begin{equation}
[j_{S, 0}, \kappa_S] = \left(\begin{array}{cc}  g\sigma^{1}C +
\bar{g}B\sigma^{1} & 2fB + g(\sigma^{1}D - A\sigma^{1}) \\ 2fC +
\bar{g}(D\sigma^{1} - \sigma^{1}A) & -\bar{g}\sigma^{1}B -
gC\sigma^{1} \end{array}\right)\;.
\end{equation}
In order for this to vanish, we need $C = B = 0$ and  $\sigma^{1}D -
A\sigma^{1} = D\sigma^{1} - \sigma^{1}A = 0$.  In turn, if we write
$A = a^{0}I + a^{a}\sigma^{a}$ and $D = d^{0}I + d^{a}\sigma^{a}$,
then this gives us
\begin{equation}
a^{0} = d^{0}, \ \ \ a^{1} = d^{1}, \ \ \ a^{2} = -d^{2}, \ \ \ a^{3} = -d^{3}.
\end{equation}

Furthermore, we find that
\begin{eqnarray}
\{j_{S, 1}, \kappa_{S}\} & = &
\end{eqnarray}
\begin{displaymath}
 \left(\begin{array}{cc} 2a^{0}\alpha^{a}\sigma^{a}
  + 2\alpha^{a}a^{a}I &  2a^{0}(\beta^{a}\sigma^{a} + \beta^{0}I)   \\
  &+ 2(\beta^{0}a^{1} - i\beta^{2}a^{3} + i\beta^{3}a^{2})\sigma^{1}  \\
2a^{0}(\gamma^{a}\sigma^{a} + \gamma^{0}I) \\+ 2(\gamma^{0}a^{1} +
i\gamma^{2}a^{3} - i\gamma^{3}a^{2})\sigma^{1} &
2d^{0}\delta^{a}\sigma^{a} + 2\delta^{a}d^{a}I\end{array}\right)\;.
 \end{displaymath}
This can be made to vanish if we set $a^{0} = 0$ and have
\begin{equation}
a^{1}\cos\psi\sin\xi - a^{2}\cos\psi\cos\xi - a^{3}\sin\psi = 0\;.
\end{equation}
The problem is that the  variables $\psi$ and $\xi$ are not necessarily constant in time because they express the direction the string moves off in as it leaves the D-brane.  Thus there is no way to define a non-trivial $\kappa_{S}$ that is independent of time.

If we were to limit down to an $SU(2)$ sector, this solves the problem by restricting the direction that the string can move off in as it leaves the D-brane: specifically, if $\psi = 0$ then $a^{1} = a^{2} = 0$ is a valid solution. Thus we conclude that a non-trivial $\kappa_S$ is possible for an $SU(2)$ subsector.

One can also show that there are no non-trivial solutions for $\kappa_A$. To see this, we can parameterize $g_A$ in (\ref{gmatrices}) using the Poincar\'e patch as in \cite{classint}. Namely,
\bea {\cal Z}_1 = Z_1 + i Z_2\;,
\;\;\; {\cal Z}_2 = Z_3 + i Z_4\;,\;\;\; {\cal Z}_3 = Z_0 + i Z_5\;,
\eea where\bea Z_0 &=& \frac{1}{2} \left(e^\phi + 2(x \bar x + x^+
x^-) e^\phi + e^{-\phi}\right)\;,\;\;\; Z_5 = \frac{e^\phi}{\sqrt
2}(x^+
- x^-)\;,\nonumber \\
Z_1 &=& \frac{1}{2} \left(e^\phi - 2(x \bar x + x^+ x^-) e^\phi -
e^{-\phi}\right)\;,\;\;\; Z_2 = \frac{e^\phi}{\sqrt 2}(x^+ + x^-)\;,
\nonumber \\
Z_3 &=& \frac{e^\phi}{\sqrt 2} (x + \bar x)\;,\;\;\; Z_4 = -i
\frac{e^\phi}{\sqrt 2} (x - \bar x)\;, \eea and, \bea x^\pm  =
\frac{1}{\sqrt 2} (x^3 \pm x^0)\;,\;\;\; x = \frac{1}{\sqrt 2} (x^1
+ i x^2)\;,\;\;\; \bar x = \frac{1}{\sqrt 2} (x^1 - i x^2)\;.\eea
The $AdS_5$ metric in these coordinates is given by, \be ds^2 = e^{2
\phi} \eta_{\mu \nu} dx^\mu dx^\nu + d\phi^2\;,\ee
so that the D5-brane is located at $x_{3} = 0$, and all other coordinates have Neumann boundary conditions.

In this case we find that
\begin{equation}
j_{A, a} = -g_{A}^{-1}\partial_{a}g_{A} = \left(\begin{array}{cc}
\frac{-1}{2}\partial_{a}\phi & 0 \\ -ie^{\phi}\partial_{a}(x_{0}I +
x_{1}\sigma_{3} + x_{2}\sigma_{1}+ x_{3}\sigma_{2}) &
\frac{1}{2}\partial_{a}\phi \end{array}\right)\;,
\end{equation}
so that at the boundary of the open string we have
\begin{equation}
j_{A, 1} = \left(\begin{array}{cc} 0 & 0 \\ -ix_{3}'e^{\phi} & 0
\end{array}\right)\;,
\end{equation}
and
\begin{equation}
j_{A, 0}  = \left(\begin{array}{cc} -\frac{1}{2}\dot{\phi}I & 0 \\
-ie^{\phi}(\dot{x}_{0}I + \dot{x}_{1}\sigma_{3} +
\dot{x}_{2}\sigma_{1}) & \frac{1}{2}\dot{\phi}I
\end{array}\right)\;.
\end{equation}
Now, if we need to satisfy $\{\kappa_{A}, j_{A, 1}\} = [\kappa_{A}, j_{A, 0}] = 0$ for some
\begin{equation}
\kappa_{A} = \left(\begin{array}{cc} A & B \\ C & D
\end{array}\right)\;,
\end{equation}
then it is straightforward to verify that, barring any further restrictions on the $AdS$ coordinates, we need the conditions
\begin{equation}
A = D, \ \ \ B = C = 0, \ \ \ \{A, \sigma^2\} = 0, \ \ \ [A,
\sigma^{3}] = 0, \ \ \ [A, \sigma^{1}] = 0
\end{equation}
which can not all be satisfied.  It is therefore not possible to
generate a working $\kappa_{A}$ or a working $\kappa_{S}$.  One
should note that were it possible to generate one, but not the
other, we would be able to produce an infinite set of non-local
charges by setting the other part of the reflection matrix to zero.
However, this would result in charges that were completely
independent of motion either on the $AdS_{5}$ or on the $S^{5}$;
these would not suggest integrability because they would not be
complete.  For example, we would not be able to use these charges to
generate all classical solutions using the algebraic curve method.

Again, if we restrict ourselves to the $SU(2)$ sector, the string will be located at the origin of $AdS$ in global coordinates. This is the same boundary condition as with the MGG. Therefore, using the global coordinates, we have the solution $\kappa_A \propto \text{diag}(1,1,-1,-1)$.

Now, in the gauge theory the $SU(2)$ sector is closed to all orders, and we know that it is integrable at one loop.  This suggests that possibly the open string in an $SU(2)$ sector attached to these D-branes might satisfy exact integrability. As mentioned earlier, the gauge theory result of integrability was independent of the number of D5-branes used.  Therefore, a possible next step would be to examine the $SU(2)$ sector of these open strings away from the probe limit, where they would more in a more complicated background.  What is interesting is that in the full bosonic sector, in contrast with the results for the giant gravitons, here we have a disagreement between the results of the one loop gauge theory calculations and the classical string results. The one loop gauge theory calculation indicated integrability in the full bosonic sector \cite{DM}.  Based on the large $\lambda$ result from the string theory, it seems unlikely this result would extend to higher loops.

To close this section, let us check that the conserved charges are in involution for the $SU(2)$ sector. For this, we note that the problem is exactly the same as with the MGG. We can use the embedding (\ref{embedding}), but now with $\phi' = 0$ at the boundaries. However, this change does not matter since, in these coordinates, the D5-brane is at $\theta_0 = 0$. Therefore, we get exactly the same reflection matrix $\kappa \propto \sigma^3$ which we know satisfy the constraints (\ref{conv}). We conclude that the conserved charges for the $SU(2)$ sector of these open stings are also in involution.

\section{Bosonic Sector for Strings ending on D7-branes}

In this section, we consider the case of a D7-brane that wraps an $AdS_5 \times S^3$ spacetime. The holographic dual is ${\cal N} = 4$, $SU(N)$ SYM theory in which we add one ${\cal N} = 2$ hypermultiplet of fundamental matter.  The integrability of this system was studied  in \cite{Erler}.  Conformal symmetry is only present in the strict large $N$ limit, which corresponds to the probe brane limit.  As in the case of the D5-brane, this is the limit that we study in this paper.

The one-loop open spin chain describing the $SO(6)$ scalar fluctuations of the D7-brane was shown to be integrable in \cite{Erler}. For the corresponding  classical open strings, we now show that the full bosonic sector is also integrable.

As before, we can separate the reflection matrix in the $AdS$ part ($\kappa_A$) and the $S^5$ part ($\kappa_S$). Since the D7-brane fills the $AdS_5$ space, all of the coordinates in these directions have Neumann boundary conditions. Therefore, we have that
\be j_{A, 1} = - g_A^{-1}
\partial_1 g_A = 0\;,\ee
at the boundaries of the open string. The condition (\ref{refcg}) becomes simply $[j_{A,0} , \kappa] = 0$. Since we do not have any further restrictions for $j_{A,0}$, we may choose $\kappa \propto \bf{1}$. On the other hand, $\kappa_S$ will need to obey similar conditions as for the MGG.  Using the angular coordinates established in section 5, we locate the brane at $\theta = 0$, with $\Omega_{2}$ and $\Omega_{3}$ satisfying Neumann boundary conditions.  Following through, we get the same block structure for the matrices $j_{A, 0}$ and $j_{A, 1}$ as for the MGG, and we thus have the same solution $\kappa_S \propto \text{diag}(1,1,-1,-1)$.  Therefore, in this case we have found that the classical open strings are integrable.  This suggests that the underlying spin chain on the gauge theory side might be integrable to all loops.

Finally, let us mention that if we restrict the string to the $SU(2)$ sector, we get exactly the same problem as with the MGG.  Thus, the charges will also be in involution.

\section{Discusssion}
In this paper, we have developed a technique to construct  non-local conserved charges for classical open strings on coset spaces. The procedure was adapted from existing techniques used for 1+1 field theories on the half line. The procedure involved the introduction of suitable ``reflection matrices" into the classical transfer matrix. These reflection matrices needed to obey certain conditions in order for the charges to be conserved, and in involution.

We studied the bosonic sector of open strings on $AdS_5\times S^5$,
which is given by the coset $\frac{SO(4,2)}{SO(5,1)} \times
\frac{SO(6)}{SO(5)}$. Boundary conditions corresponding to Giant
Gravitons, D5-branes and D7-branes were studied. We found that
strings ending on ``Maximal" Giant Gravitons and D7-branes were
integrable. In contrast, we found that we could not construct
non-local charges for open strings on non-maximal Giant Gravitons.
This is unless we imposed extra boundary conditions which were very
restrictive. Some of these conditions agree with expectations from
the gauge theory \cite{bcv}.

For D5-branes, we found that we could only construct the conserved charges for the $SU(2)$ sector. This is very interesting, since the full bosonic sector seems to be integrable at one-loop in the gauge theory \cite{DM}. Therefore, it seems that integrability is broken at some higher order in $\lambda$, or possibly non-perturbatively, for the full open spin chain.  It would be interesting to see this breakdown of integrability by a direct gauge theory calculation. This kind of behavior is not unheard of. Breakdown of perturbative integrability was observed some time ago in the Plane Wave Matrix Model \cite{breakdown}. Nevertheless, we remind the reader that we have only given {\it evidence} for non-integrability. It is possible that the construction of classical conserved charges is still possible using a different technique.

For the case of the Maximal Giant Graviton, it has been argued that integrability might be broken at higher loops \cite{2loop}. This was based on an apparent failure  to construct a Perturbative Asymptotic Bethe Ansatz (PABA) for the $SU(2)$ sector. More specifically, the problem arose in trying to explicitly construct the two-particle wavefunction at two loops.

On the other hand, our results indicate that integrability is present at vert strong coupling for the full bosonic sector.  It would be odd if integrability were only realized at very weak and very strong coupling, but not for general values of $\lambda$.  Therefore, we believe that the spin chain describing open strings on Maximal Giant Gravitons might indeed be integrable at higher loops.  Our results suggest that more effort should be put into understanding the role of the PABA when boundary conditions are included, and into looking for solutions to the problems proposed in \cite{2loop}.

Let us mention a possible route to make some progress in this
direction. It has been recently shown in \cite{BetheSigma1,
BetheSigma2}, that one can match the low-energy limit of the
S-matrix that enters the PABA, with the quantum S-matrix of the
Landau-Lifshitz reduced string action. It is then very desirable to
extend these techniques for open strings. In particular, one would
like to match the PABA reflection matrices studied in
\cite{2loop,2loopok} with the quantum reflection matrices that come
directly from the reduced sigma model. This can be accomplished by
using the techniques discussed in \cite{Kim, Bajnok}. This is
especially interesting since, in contrast with the bulk S-matrix, we
have seen that the reflection matrices have a {\it classical} limit.
Therefore, they allow for a more direct check of the AdS CFT
correspondence. In particular, we have seen that, when restricted to
the $SU(2)$ sector, the classical reflection matrix has a universal
form $\kappa \propto \sigma^3$. It would be interesting to
understand this from the underlying spin chains.

Finally, it would  also be interesting to apply our techniques to
the full $PSU(2,2|4)$ sector for the MGG and D7-brane systems. The
main complication with this calculation is the need for a gauge
choice in the world sheet.  One approach that has been used in the
past for explicit construction of the matrix $g$, is to use a
light-cone gauge in the Poincar\'e patch of $AdS_{5}$ \cite{gauge}.
This in particular is unsuitable for the Maximal Giant Graviton,
since it is located at the center of $AdS_{5}$. However, once a
gauge choice is made, one can study the boundary conditions on the
fermions using the techniques of \cite{west}. Another possible route
would be to expand around a plane-wave background. We think that,
technical difficulties aside, the full open superstring attached to
the MGG is classically integrable.

\acknowledgments

We would like to thank David Berenstein,  Joe Polchinski, and Oleg
Lunin for useful discussions.  The work of S.E.V. was supported, in
part, by a National Science Foundation Graduate Fellowship.  The
work of N.M. was partially supported by National Science Foundation
grant PHY00-98395, partially supported by Department of Energy grant
580093, and partially supported by the Enrico Fermi Institute
endowment 750573.  Any opinions, findings, conclusions or
recommendations expressed in this material are those of the authors
and do not necessarily reflect the views of the National Science
Foundation, the Department of Energy, or the Enrico Fermi Institute.

\end{document}